\documentstyle[fleqn,12pt]{article}

\def\pj{\hspace{-.26cm}}
\def\fpj{\hspace{-.7cm}}
\def\thalf{{\textstyle{\frac{1}{2}}}}

\def\tquar{{\textstyle{\frac{1}{4}}}}

\newcommand{\vm}[1]{\mbox{\bf#1}}
\newcommand{\vmg}[1]{\mbox{\boldmath$#1$}}
\begin{document}
\title{An Effective Lagrangian with Broken Scale and Chiral Symmetry
Applied to Nuclear Matter and Finite Nuclei}
\author{Erik K. Heide, Serge Rudaz and Paul J. Ellis\\[4mm]
{\small School of Physics and Astronomy}\\[-2mm] {\small University of
Minnesota, Minneapolis, MN 55455}}
\date{~}
\maketitle
\vskip-1cm
\begin{center}
{\small\bf Abstract}
\end{center}
\vskip-.25cm
{\small We study nuclear matter and finite nuclei with a chiral Lagrangian
which generalizes the linear $\sigma$ model and also accounts for the QCD
trace anomaly by means of terms which involve the $\sigma$ and $\vmg{\pi}$
fields as well as the glueball field $\phi$. The form of the scale invariant
term leading to an omega meson mass, after symmetry breaking, could involve
coupling to $\phi^2$ or $\sigma^2$ or
some linear combination thereof. In fact an $\omega_{\mu}\omega^{\mu}\phi^2$
form is strongly favored by the bulk properties of nuclei, which also rather
strongly constrain the other parameters. A reasonable description of the
closed shell nuclei oxygen, calcium and lead can be achieved, although the
spin-orbit splittings are somewhat smaller than the experimental values.}
\thispagestyle{empty}
\vskip-21.5cm
\hfill NUC-MINN-93/21-T\\
\phantom{111}\hfill UMN-TH-1210/93\\
\phantom{111}\hfill JULY 1993\\
\newpage
\section{Introduction}

We are still far from being able to transform the  Lagrangian of
quantum chromodynamics (QCD)
into an effective Lagrangian involving mesons and baryons which could
be used for nuclear matter or finite nuclei. Nevertheless it is possible
to incorporate the broken global chiral and scale symmetries of QCD into an
effective Lagrangian.
In particular the notion of (broken) scale invariance has been modeled
by a scalar glueball potential \cite{mig}. An early attempt to introduce
this into an appropriately modified Walecka-type Lagrangian was made in ref.
\cite{eko1}. An alternative approach is to include the glueball potential
in the Lagrangian of the linear sigma model (supplemented by a contribution
from the vector-isoscalar omega field) \cite{glu1,rho}.
The latter is particularly attractive since it incorporates (spontaneously
broken) chiral symmetry, another feature suggested by QCD. A common problem
associated with these models is that the compression modulus for
equilibrium nuclear matter is at least a factor of two larger than
current estimates in the range 200--300 MeV. Of course one can arbitarily
correct this by adding terms to the Lagrangian, for instance
$\sigma^3$ and  $\sigma^4$, and choosing the parameters to produce the
desired result \cite{rod}. We have recently suggested \cite{glu2}
a more satisfactory approach, however, where the form of the
potential which breaks scale invariance is modified
in a reasonable way to include a contribution from the $\sigma$-field as
well as the glueball field, $\phi.$ This was found to yield $K\approx390$ MeV,
which, while still on the large side, is decidedly more reasonable.

There still remains uncertainty concerning the form of the
omega mass term, $\thalf m_{\omega}^2\omega_{\mu}\omega^{\mu}$, where the
explicitly dimensionful parameter $m_{\omega}^2$ must be replaced by the
square of a scalar field (multiplied by a dimensionless coupling constant)
in order to satisfy the dictates of scale invariance.
The question arises as to whether the coupling be to $\sigma^2$
or $\phi^2$ or some linear combination of the two. In
refs. \cite{glu1,glu2} the first choice was made. One of the purposes of the
present paper is to examine this choice in nuclear matter and finite nuclei;
we shall find that the properties of finite nuclei are sensitive to this
question.
A more basic purpose of this investigation is to determine whether our
effective Lagrangian, which
saturates nuclear matter in the mean field approximation, is able to give a
sensible account of finite nuclei. This is particularly pressing since
Furnstahl and Serot \cite{fern} have recently suggested that chiral models are
inherently unable to give a satisfactory description of nuclei; this work was
based on the approach of Boguta \cite{bog} and did
not include the glueball field, so there was no attempt to incorporate
broken scale invariance.

The effective Lagrangian that we employ, together with the necessary
formalism for finite nuclei and nuclear matter is given in sec. 2. Our
results are displayed in sec. 3, first for nuclear matter and then for the
closed shell nuclei $^{16}$O, $^{40}$Ca and $^{208}$Pb. Our conclusions are
given in sec. 4. It is of interest to compare the low density expansion to
that of the standard Walecka model \cite{sew}, this we discuss briefly in
the Appendix. This suggests a simple approximation in which the glueball
field is frozen at its vacuum value and the formalism is outlined in the
Appendix. Examples are given in the text which show that the frozen glueball
model reproduces our complete results quite accurately.

\section{Theory}

We write the total effective Lagrangian in the form
\begin{equation}
{\cal L_{\rm eff}}={\cal L}_0-V_G\;,
\end{equation}
where in this schematic equation we have separated out the scale invariant
part, ${\cal L}_0$, from the potential $V_G$ which induces the breaking of
scale and chiral invariance. (The effects of explicit chiral symmetry breaking
arising from the non-vanishing of current quark masses in the QCD Lagrangian
are neglected.)
We write ${\cal L}_0$ in terms of the chiral-invariant combination of sigma and
pi fields, $\sigma$ and $\vmg{\pi}$, the glueball field $\phi$,
the field of the omega vector meson $\omega_{\mu}$ and, since we are
interested in finite nuclei, the vector-isovector rho field ${\bf b}_{\mu}$
and the Maxwell field $A_{\mu}$. Specifically
\begin{eqnarray}
&&\fpj{\cal L}_0=\thalf\partial_{\mu}\sigma\partial^{\mu}
\sigma+\thalf\partial_{\mu}\vmg{\pi}\cdot\partial^{\mu}\vmg{\pi}
+\thalf\partial_{\mu}\phi\partial^{\mu}\phi-\textstyle{\frac{1}{4}}
F_{\mu\nu}F^{\mu\nu} -\tquar \vm{B}_{\mu\nu}\cdot\vm{B}^{\mu\nu}
\nonumber\\
&&\quad-\textstyle{\frac{1}{4}}f_{\mu\nu}f^{\mu\nu}
+\thalf \omega_{\mu}\omega^{\mu}[G_{\omega\sigma}(\sigma^2+\vmg{\pi}^2)
+G_{\omega\phi}\phi^2]
+\thalf G_{\rho}\phi^2\vm{b}_{\mu}\cdot\vm{b}^{\mu}\nonumber\\
&&\fpj+\bar N\bigl[\gamma^{\mu}(i\partial_{\mu}-g_{\omega}\omega_{\mu}
-\thalf g_{\rho}\vm{b}_{\mu}\cdot{\vmg\tau}-\thalf e(1+\tau_3)A_{\mu})
-g(\sigma+i\vmg{\pi}\cdot\vmg{\tau}\gamma_5)\bigr]N\;.\nonumber\\
&&
\end{eqnarray}
Here the field strength tensors are defined in the usual way
$F_{\mu\nu}=\partial_{\mu}\omega_{\nu}-\partial_{\nu}\omega_{\mu}$,
$\vm{B}_{\mu\nu}=\partial_{\mu}\vm{b}_{\nu}-\partial_{\nu}\vm{b}_{\mu}$ and
$f_{\mu\nu}=\partial_{\mu}A_{\nu}-\partial_{\nu}A_{\mu}$. As we have remarked
in the introduction, the form of the scale invariant $\omega$ mass term is not
{\it a priori} obvious, so we have chosen a linear
combination $\sigma^2$ and $\phi^2$ fields (of course more exotic choices are
possible) and we shall examine the effect
of varying the coefficients. In the corresponding case for the $\rho$ meson
we have simply chosen $\phi^2$, we shall comment further upon this later.

In order to obtain the scale breaking term $V_G$ in the Lagrangian, we recall
that the divergence of the scale current in QCD is given by the trace
of the ``improved" energy-momentum tensor and so the scalar potential
$V_G(\phi,\sigma,\vmg{\pi})$ is chosen to
reproduce, via Noether's theorem, the effective trace anomaly \cite{mig,gomm}
\begin{equation}
\theta^{\mu}_{\mu}=4V_G(\Phi_i)-\sum_i\Phi_i\frac{\partial V_G}{\partial\Phi_i}
=4\epsilon_{\rm vac}\left(\frac{\phi}{\phi_0}\right)^{\!4}\;,
\end{equation}
where $\Phi_i$ runs over the scalar fields \{$\phi,\sigma,\vmg{\pi}$\} and
$\epsilon_{\rm vac}$ is the vacuum energy.
The proportionality $\theta^{\mu}_{\mu}\propto\phi^4$ is suggested by
the form of the QCD trace anomaly \cite{crew},
\begin{equation}
\theta^{\mu}_{\mu}(x) =\frac{\beta(g)}{2g}F^a_{\mu\nu}(x)F^{a\mu\nu}(x)\;,
\end{equation}
where $F^a_{\mu\nu}(x)$ is the gluon field strength tensor and $\beta(g)$ is
the usual QCD beta function. Eq. (3) involves a purely gluonic, color-singlet,
scalar, dimension-four operator, and also allows for the recovery of the low
energy theorems which follow from broken scale invariance \cite{itep}.
Then in ref. \cite{glu2} we suggested the form
\begin{eqnarray}
V_G(\phi,\sigma,\vmg{\pi})=B\phi^4\left(\ln\frac{\phi}{\phi_0}-\tquar\right)
&\pj-&\pj\thalf B\delta
\phi^4\ln\frac{\sigma^2+\vmg{\pi}^2}{\sigma_0^2}\nonumber\\
&&\quad+\thalf B\delta \zeta^2\phi^2\!\!\left(\sigma^2+\vmg{\pi}^2
-\thalf\frac{\phi^2}{\zeta^2}\right)\;,
\end{eqnarray}
where $\zeta=\frac{\phi_0}{\sigma_0}$ and the subscript 0 indicates
the vacuum value.
Here the logarithmic terms
contribute to the trace anomaly and are such that eq. (3) is
satisfied with $\epsilon_{\rm vac}=-\tquar B\phi^4_0(1-\delta)$. This
requirement uniquely specifies the second term on the right in eq. (5). The
third term is needed to ensure that in the vacuum
$\phi=\phi_0$, $\sigma=\sigma_0$ and $\vmg{\pi}=0$. To
retain the physically necessary feature that $\epsilon_{\rm vac}<0$, given
that $B>0$, we require that $\delta<1$. Provided $0<\delta<1$,
$V_G\rightarrow\infty$ for $\sigma$ or $\phi\rightarrow\infty$ as is physically
sensible. For further insight regarding the parameter $\delta$, we recall
that the QCD beta function with
$N_c$ colors and $n_f$ flavors is given at the one loop level by
\begin{equation}
\beta(g)=-\frac{11N_cg^3}{48\pi^2}\left(1-\frac{2n_f}{11N_c}\right)
+{\cal O}(g^5)\;,
\end{equation}
where the first number in parentheses arises from the (antiscreening)
self-interaction of the gluons and the second, proportional to $n_f$, is the
(screening) contribution of quark pairs. Recalling eq. (4), a value of
$\delta=4/33$ is suggested for the present case with $n_f=2$ and $N_c=3$.
Since one cannot rely on the one-loop estimate for $\beta(g)$, we shall examine
the effect of modifying the value of $\delta$.

The careful reader will have noticed that ${\cal L}_{\rm eff}$, given by eqs.
(2) and (5), does not contain the standard potential term of
the linear sigma model which we have previously written \cite{glu1,glu2}
in scale invariant form as
$\tquar\lambda\left(\sigma^2+\vmg{\pi}^2-\frac{\phi^2}{\zeta^2}\right)^{\!2}$.
We found that very small
values of $\lambda$ were favored by the predicted compression modulus in
nuclear matter. We have
also found that departures from small values yield binding energies of nuclei
which are much too low. Therefore in this presentation we make the welcome
simplification of setting $\lambda=0$, thus eliminating this term. Note that
while the term is present in the standard linear sigma model, it is {\it not}
necessary for the breaking of chiral invariance. This can be carried out
just as well by $V_G$ in eq. (5).

It is economical to write
the sigma and glueball fields in terms of their ratios to the vacuum values,
{\it viz.}
\begin{equation}
\chi(r)=\frac{\phi(r)}{\phi_0}\qquad,\qquad\nu(r)=\frac{\sigma(r)}{\sigma_0}\;.
\end{equation}
For $r\rightarrow\infty$ we obtain the vacuum values $\chi=\nu=1$. Now, in
the vacuum, we require that the rho and the omega masses
take their physical values so that we can write the mass terms in the form
\begin{equation}
\thalf m_{\rho}^2\chi^2\vm{b}_{\mu}\cdot\vm{b}^{\mu}\quad{\rm and}\quad
\thalf m_{\omega}^2\left[R_{\omega}\nu^2+(1-R_{\omega})\chi^2\right]
\omega_{\mu}\omega^{\mu}\;.
\end{equation}
Thus, for the omega, taking $R_{\omega}$ to be 1 gives a pure $\nu$ coupling,
whereas the value 0 yields a pure $\chi$ coupling. The effective mass of the
nucleon, in the usual way , is $M^*(r)=M\nu(r)$.
The field equations are obtained from Lagrange's equations and of course
$\vmg{\pi}=0$ in the mean field approximation. We specialise
to the time independent, spherically symmetric case. The Dirac equation for the
nucleon and the equation for the photon field are of the form given
by Horowitz and Serot \cite{horse} and do not need to be repeated here.
As usual for the vector fields only the time-like components survive and also
just the isovector $z$-component for the  $\rho$ field.
The equations for the $\phi$, $\sigma$, $\omega$ and $\rho$ can be written
\begin{eqnarray}
&&\fpj\phi_0^2D\chi-2B_0(2-\delta)\chi+2B_0\delta\nu=
4B_0[\chi^3(\ln\chi-\delta\ln\nu)-\chi+\delta\nu]\nonumber\\
&&\quad+B_0\delta[\chi(\nu^2-\chi^2)+2(\chi-\nu)]
-m_{\omega}^2(1-R_{\omega})\omega_0^2\chi-m_{\rho}^2b_0^2\chi\;,\nonumber\\
&&\fpj\sigma_0^2D\nu-2B_0\delta\nu+2B_0\delta\chi=M\rho_s
-B_0\delta\left[\frac{\chi^4}{\nu}-\chi^2\nu+2(\nu-\chi)\right]\nonumber\\
&&\hspace{7.5cm}-m_{\omega}^2R_{\omega}\omega_0^2\nu\;,\nonumber\\
&&\fpj D\omega_0-m_{\omega}^2\omega_0=-g_{\omega}\rho_B+m_{\omega}^2
[R_{\omega}(\nu^2-1)+(1-R_{\omega})(\chi^2-1)]\omega_0\;,\nonumber\\
&&\fpj Db_0-m_{\rho}^2b_0=-g_{\rho}\rho_3-m_{\rho}^2(1-\chi^2)b_0\;,
\end{eqnarray}
where the densities $\rho_s=\langle\bar{N}N\rangle$,
$\rho_B=\langle\bar{N}\gamma^0N\rangle$ and
$\rho_3=\thalf\langle\bar{N}\gamma^0\tau_0N\rangle$ can be expressed in terms
of the components of the nucleon Dirac spinors in the usual way \cite{horse}.
In eq. (9) we have made the definitions $D\equiv\frac{d^2}{dr^2}
+\frac{2}{r}\frac{d}{dr}$ and $B_0\equiv B\phi_0^4$. The terms linear in the
fields, {\it i.e.}, the kinetic energy and mass terms, have been separated out
on the left of these equations. The equations of ref. \cite{glu2} for
nuclear matter are regained by setting the derivatives to zero so that
the results are independent of $\phi_0$ and $\sigma_0$, however these
quantities are needed for finite nuclei. They are also needed to obtain the
vacuum scalar masses and we note that the mass matrix is not diagonal. In
order to solve the equations (9) by an iterative Green's function technique
\cite{fern,horse} it is necessary to go to a representation which is diagonal
in the limit that $r\rightarrow\infty$. We solve in terms of the quantities
$(1-\chi)$  and $(1-\nu)$ which go to zero in the same limit.

The energy-momentum tensor can be used to obtain the total energy of the
system in the standard way \cite{sew}. Subtracting constants so that the
energy is measured relative to the vacuum, we obtain
\begin{eqnarray}
E&\pj=&\pj\sum_{\alpha}^{\rm occ}\epsilon_{\alpha}(2j_{\alpha}+1)
-2\pi\int_0^{\infty}dr r^2\bigg\{M\nu\rho_s+g_{\omega}\omega_0\rho_B
+g_{\rho}b_0\rho_3\nonumber\\
&&+2B_0[\chi^4(\ln\chi-\delta\ln\nu+\tquar)-\tquar]+\thalf B_0\delta
[\chi^2(2\nu^2-3\chi^2)+1]\nonumber\\
&&\qquad\qquad-m_{\omega}^2[R_{\omega}\nu^2+(1-R_{\omega})\chi^2]\omega_0^2
-m_{\rho}^2\chi^2b_0^2\bigg\}\;.
\end{eqnarray}
In the first term on the right the $\epsilon_{\alpha}$ are the Dirac single
particle energies and $j_{\alpha}$ is the total angular momentum of the single
particle state. In nuclear matter this term becomes
$4\sum_{\vm{k}}\left(g_{\omega}\omega_0+\sqrt{k^2+M^{*2}}\right)$. Making this
replacement and using eq. (9) the expression for the energy density of
nuclear matter given in ref. \cite{glu2} can be obtained.

\section{Results}
\subsection{Nuclear Matter}

In nuclear matter we have four parameters to consider. We will choose
$B_0$ and $C_{\omega}^2\equiv\frac{g_{\omega}^2M^2}{m_{\omega}^2}$ to fit
the saturation properties. The binding energy/nucleon we take
to be 16 MeV. In refs. \cite{glu1,glu2} we took a saturation density of
0.16 fm$^{-3}$, but this yields central densities in lead which are too high.
We have therefore used 0.148 fm$^{-3}$ ($k_F=1.3$ fm$^{-1}$), which is the
value favored by Serot and coworkers in mean field calculations
\cite{horse,fern}. Having fixed these two parameters, we then have the
parameter of the $\omega$ mass term, $R_{\omega}$, and the coefficient of the
quark contribution to the trace anomaly, $\delta$, which can be varied.

We show in table 1 several parameter sets that we have considered (the
designation F here indicates the frozen glueball model where $\chi$ is kept
equal to 1; see discussion in the Appendix). One might hope that the parameters
would show a reasonable correspondence with the independently determined
quantities, although precise agreement would not be expected for
an effective theory such as this.
In this sense the values for $C^2_{\omega}$ in table 1 are in acceptable
agreement with the value $103\pm36$ deduced  from data on the $\omega$ coupling
constant \cite{oades} and the known mass.
Turning to $|\epsilon_{\rm vac}|$, QCD sum rule studies suggest \cite{svz} a
value of $(240\ {\rm MeV})^4$ which would not favor the extreme variations
of $\delta$ (sets IV and VII), although bag model estimates \cite{bag} range as
low as (146 MeV)$^4$. The values of $C_{\rho}^2$, given in table 1 for
later use, have been obtained by fitting to a symmetry energy of 35 MeV.
Since exchange effects are
known to make a significant contribution to the symmetry energy we shall not
make a direct comparison to experiment.

Turning to the derived quantities, nuclei generally
favor effective masses below 0.7 \cite{fern}, which is the case for sets VI
and VII. However, here the compression moduli, $K$, are in excess of 400 MeV.
Pearson \cite{pear} has made the point that the leptodermous
expansion, which is one method used to fit the data, permits values of $K$
up to about 400 MeV with little difference in the $\chi^2$ fits. Defining
the third derivative of the binding energy per particle as
$S=k^3_F\frac{d^3}{dk_F^3}\left(\frac{E}{A}\right)$ (evaluated at equilibrium),
the data indicate a linear relation between $S/K$ and $K$
(see ref. \cite{skpap}). We have listed values of $S/K$ in table 1.
Cases VI and VII are well off the allowed error band, but the remaining cases
are quite close to, although slightly off, the band.

The parameter sets designated with an F in table 1 correspond to the frozen
glueball model ($\chi=1$) and we see that they give a close correspondence
with the complete results. The point is further made in figs. 1 and 2 where the
agreement is seen to be particularly good at low density, as we argue
in the Appendix. The lower panel of fig. 2 shows that in the complete model
$\chi$ remains very close to unity over the whole region of densities
considered. We remark that the curves for $\nu$ will increase linearly with
$k_F$ for sufficiently large density (a discussion of the high density
behavior was given in ref. \cite{glu1}), although it is debatable whether this
theory is applicable in that region.

\subsection{Finite Nuclei}

When we turn to finite nuclei two extra parameters are involved, namely the
vacuum values of the scalar fields, $\phi_0$ and $\sigma_0$. Thus we have a
total of four free parameters. Let us first eliminate two of these. We
shall discuss changes relative to parameter set I of table 1 with
$\sigma_0 =110$ MeV, which will turn out to be our preferred parameter set.
 First consider reducing $\delta$ by using sets IV
and V of table 1 (and reasonable values for the parameters not specified). We
find little sensitivity in the predicted properties of nuclei. However, if we
increase
$\delta$ using sets VI and VII, for which $K$ is large, we find a significant
reduction in the binding energy, particularly for set VII ($^{40}$Ca is less
bound by 2$\tquar$ MeV, for example). We conclude that a value of $\delta$ in
the neighborhood of 4/33 is reasonable and fix on this value, although nuclei
are not sensitive to the precise figure. Next consider the ratio of the
vacuum scalar fields,
$\zeta=\phi_0/\sigma_0$. Using set I we have varied $\zeta$ in the range
0.7--2.1, which corresponds to varying the higher scalar mass, $m_>$, between
3 and 1 GeV. Very little change is observed in the nuclear properties.
The salient point is that the
mixing between the glueball and the sigma is small, $\le3$\%, and it is
the sigma which directly couples to nucleons. Henceforth we choose $\zeta$
such that $m_>=1.5$ GeV because this seems reasonable in view of QCD sum rule
estimates for a scalar glueball mass in the range 1--2 GeV \cite{shif}.
The values of $\zeta$ are listed in table 2.

We are then left to consider the variation of $R_{\omega}$ and $\sigma_0$:
a few comments on the expected magnitude of $\sigma_0$ are in order here.
In the original Gell-Mann-L\'evy linear sigma model \cite{linsig}, a
calculation of the axial vector current matrix element responsible for pion
decay led to the identification $\sigma_0=f_{\pi}=93$ MeV. In a more general
chiral effective Lagrangian incorporating vector mesons (ref. \cite{gas} for
an early review) this identification no longer follows due to the
necessity of including the $a_1$ axial vector meson, together with the $\rho$.
Briefly, cross-terms of the form $\vm{a}_1^{\mu}\cdot\partial_{\mu}\vmg{\pi}$
appear in the full Lagrangian which are eliminated by the replacement
$\vm{a}_1^{\mu}\rightarrow\vm{a}_1^{\mu}+\xi\partial^{\mu}\vmg{\pi}$, with an
appropriate parameter $\xi$. This, in turn, leads to a change in the
coefficient of the pion kinetic term,
$\partial_{\mu}\vmg{\pi} \cdot\partial^{\mu}\vmg{\pi} $, and a rescaling of
the pion field is necessary to bring this to canonical form. One defines a
renormalized field $\vmg{\pi}_R=Z_{\pi}^{-\frac{1}{2}}\vmg{\pi}$. Now a
calculation of the full axial current yields
$\sigma_0=Z_{\pi}^{\frac{1}{2}}f_{\pi}$. The
precise form of $Z_{\pi}$ is model dependent. In the simplest model, which does
not include the physics of broken scale invariance one finds \cite{gas}, using
the KSRF relation, that $Z_{\pi}^{\frac{1}{2}}=m_{a_1}/m_{\rho}=\sqrt{2}$.
We do not
expect either this precise functional form or value to be maintained in more
realistic models, but we do anticipate $Z_{\pi}^{\frac{1}{2}}>1$. Accordingly,
$\sigma_0$ is kept as a free parameter, which we can presume to be larger than
93 MeV. Two
further remarks are in order. First, the additional terms involving the $a_1$
will not contribute in the mean-field approximation.
Second, from eq. (2) after symmetry breaking, the nucleon mass is given by
$M=g\sigma_0$. The physical pion-nucleon coupling is defined in terms
of the renormalized pion field to be $g_{\pi NN}=Z_{\pi}^{\frac{1}{2}}g$. So,
in fact, one still has $M=g\sigma_0=g_{\pi NN}f_{\pi}$ as the approximate form
of the Goldberger-Treiman relation.

With this in mind, we first consider variation of $\sigma_0$ keeping
$R_{\omega}=0$. The results are given in table 2, where the notation I/93,
for example, implies use of parameter set I from table 1 with $\sigma_0=93$
MeV.
The theoretical values of the binding
energy/particle include a correction for the  c.m. kinetic energy \cite{neg}.
The charge densities, and therefore the radii quoted, are corrected for the
finite size of the proton \cite{horse} and for c.m. effects \cite{tas}.
We see that the choice $\sigma_0=93$ MeV (fourth row in table 2)
yields poor binding energies, radii which are too small and,
as shown in fig. 3 for $^{40}$Ca, an  oscillatory charge density. This figure
indicates that increasing $\sigma_0$ to 110 MeV improves matters greatly.
One can view this as reducing the magnitude of the derivative of $\nu$ in
eq. (9) or as reducing $m_<$ to $\sim500$ MeV (Furnstahl and Serot \cite{fern}
stress that values in this neighborhood are to be preferred). In
fact as regards the charge radii and densities the value 120 MeV gives even
better results, however we see from table 2 that in this case the binding
energies/particle
are 1--1$\frac{3}{4}$ MeV too low. Therefore we prefer the parameter
set I/110.

Now we consider variation of $R_{\omega}$ with $\sigma_0=110$ MeV.
As $R_{\omega}$ is increased from zero to 0.5 or 1.0
we see that the binding energies become too large and shows the wrong
$A$-dependence and also the radii become too small. Table 2 also shows that
the mass
of the lighter scalar meson, $m_<$, increases sharply. Further, as we see for
$^{208}$Pb in fig. 4 the charge density begins to develop oscillations which
are not present in the data. Similar problems were noted by Furnstahl and Serot
\cite{fern} who used $R_{\omega}$ values of 0.76 or 1, although their
model differs significantly from ours. As pointed out by
Price, Shepard and McNeil \cite{shep} this is likely to signal the onset of
a situation where the ground state of nuclear matter is no longer uniform,
but exhibits periodic density fluctuations. We conclude that nuclei strongly
favor $R_{\omega}=0$.

We have also compared the case I/110 with the frozen glueball approximation,
case IF/110, in table 2. There is a close correspondence between the results,
as there is for the predicted charge densities which we illustrate for
$^{16}$O in fig. 5.

Finally we turn to the single particle energies. The levels near
the Fermi surface in lead are shown in figs. 6 and 7 for neutrons and protons
respectively. The relative positions of the
neutron and proton levels are sensitive to the treatment of the $\rho$ meson
\cite{horse}. We have used a scale-invariant $\phi^2$ coupling in the
mass term, but the results differ negligibly if a standard mass term is used.
However there is a difference if a $\sigma^2$ coupling is used because this
reduces $C_{\rho}^2$ and leads to $\sim2$ MeV less (more) binding for
protons (neutrons). Since figs. 6 and 7 clearly show that this is undesirable
we have not considered this possibility further.
The results with the sets II/110 ($R_{\omega}=0.5$) and I/93 show immediate
problems since the major shell closure is incorrectly given due to the
position of the proton $1h_{9/2}$ and the neutron $1i_{11/2}$ levels.
This problem disappears with the preferred parameter set, I/110, and the
single particle spectrum is much more reasonable, although
a larger proton $1h_{9/2}-3s_{1/2}$ splitting would be desirable. This may
reflect the spin-orbit splittings which are, on average, 65\% of the
experimental values here and also in the other nuclei we have considered.
They could be improved by reducing the effective masses in table 1 below 0.7.
This can be achieved by taking a small negative value for $R_{\omega}$,
although we find that this leads to a larger compression modulus and lower
binding energies.

\section{Conclusions}
Our primary purpose has been to see whether an effective Lagrangian which
includes the breaking of scale and chiral invariance can make sensible
predictions for finite nuclei. Finite nuclei are expected, and found, to be a
much more severe test than nuclear matter. In this endeavour the form of the
scale invariant mass term for the $\omega$ meson was found to be critical
and a pure coupling to the glueball field,
$\phi^2\omega_{\mu}\omega^{\mu}$, was required, although a small effect from
the sigma field cannot be ruled out.
A similar coupling for the $\rho$ mass term was also favored. It is
interesting that the properties of nuclei can be used to provide such strong
constraints on the form of the couplings of the scalar glueball to other
mesonic fields appearing in the chiral effective Lagrangian.

It was also found to be necessary to increase the value
of the vacuum value of the sigma field, $\sigma_0$, from the naive
expectation of 93 MeV to somewhere in the neighborhood of 110 MeV.
With these caveats the bulk properties of O, Ca and Pb were quite well
accounted for and, while the spin-orbit splittings are about 65\% of the
desired values, the basic structure of the single particle spectra was
reasonable. As a by-product we were able to define a simplified, frozen
glueball model which was able to reproduce the results of the complete model
quite accurately, but, by itself, would seem highly arbitary.

It is natural to compare this approach to the standard Walecka model
extended to include non-linear $\sigma^3$ and $\sigma^4$ terms
(as for example in refs. \cite{fern2,fern}). The agreement is better
than found here,
particularly as regards the single particle energies. Nevertheless,
it is remarkable that the present effective chiral Lagrangian, whose form is
motivated by QCD, can give an acceptable
description of both nuclear matter and finite nuclei in the mean field
approximation. It remains to be seen whether fine tuning can further
improve the agreement here. For example, a term of the form
$(\omega_\mu\omega^\mu)^2$ is the simplest among many possibilities,
consistent with scale invariance, which could be added to the Lagrangian.
We are currently investigating the effects of including such a term.

After this work was completed we received a preprint from Furnstahl and
Serot \cite{fs2} who have studied the same problem, restricting themselves
to an $\omega$ mass term of the form $\sigma^2\omega_{\mu}\omega^{\mu}$.
In agreement with our conclusions, they point out that the results for
nuclei are poor. As we have stressed, nuclei require that the coupling
be to the glueball field rather than the sigma field.

We thank G.F. Bertsch and A. Vischer for useful comments and C.J. Horowitz
for giving us a copy of his computer code for finite nuclei which was
suitably modified for present purposes.
We acknowledge partial support
from the Department of Energy under contracts No. DE-FG02-87ER40328
and DE-AC02-83ER40105. A grant for computing time from the Minnesota
Supercomputer Institute is gratefully acknowledged.

\newpage
\begin{center}
{\bf Appendix\\
Low Density Expansion and Frozen Glueball Model}
\end{center}

It is of interest to study the low density expansion of eqs. (9) and (10) in
the case of nuclear matter. Examination of eq. (9) indicates that we can
expand
\begin{eqnarray}
\nu&\pj=&\pj1+a\rho_B(1+bk_F^2)+{\cal O}(k_F^7)\;,\nonumber\\
\chi&\pj=&\pj1+c\rho_B(1+dk_F^2)+{\cal O}(k_F^7)\;,
\end{eqnarray}
where $\rho_B$ is the baryon density and $k_F$ the Fermi momentum.
Further to this order the coupling between $\omega_0$ and $\nu$ and $\chi$
can be neglected. Then the equation of motion for $\chi$ indicates that
\begin{equation}
d=b\quad,\quad\frac{c}{a}=\frac{\delta}{2-\delta}\;,
\end{equation}
The equation of motion for $\nu$ yields
\begin{equation}
a=-\frac{M(2-\delta)}{4B_0\delta(1-\delta)}\quad,\quad b=-\frac{3}{10M^2}\;.
\end{equation}
Substituting into eq. (10), we find the energy/particle
\begin{eqnarray}
\frac{E}{A}&\pj=&\pj M+\frac{3k_F^2}{10M}-\frac{3k_F^4}{56M^3}
+\frac{g_{\omega}^2\rho_B}{2m^2_{\omega}}\nonumber\\
&&\quad-\frac{M^2(2-\delta)}{8B_0\delta(1-\delta)}\rho_B
\left(1-\frac{3k_F^2}{5M^2}\right)+{\cal O}\left[\left(\frac{k_F}{M}
\right)^{\!6}\right]\;.
\end{eqnarray}
The first three terms are the energy of a Fermi gas and  the succeeding terms
are the $\omega$ and scalar meson contributions.
This is of exactly the same form as the standard Walecka model \cite{sew},
except that the scalar meson mass is replaced by
\begin{equation}
m=\left(\frac{4B_0\delta(1-\delta)}{\sigma_0^2(2-\delta)}
\right)^{\!\thalf}
= m_<\:m_>\:\frac{\phi_0}{[2B_0(2-\delta)]^{\thalf}}\simeq m_<\;.
\end{equation}
where $m_<$ ($m_>$) are the lower (upper) eigenvalues of the vacuum mass matrix
(see \cite{glu2} for an explicit discussion). The mass, $m$, is
approximately $m_<$, provided that $\phi_0/\sigma_0$ is not too large so that
the mixing of the sigma and glueball states is small.

Now the fact that $d=b$ in eq. (12), means that, to the order we are
considering, we can obtain the correct result for $\nu$ with $\chi=1$,
provided that $B_0$ is replaced by
\begin{equation}
B_0'=B_0\frac{2(1-\delta)}{2-\delta}
\end{equation}
In this approximation, we see that the scalar potential, $V_G$, takes the
 simple form
\begin{equation}
V_G(\phi,\sigma,\vmg{\pi}=0)-\epsilon_{\rm vac}=\thalf B_0\delta
(\nu^2-1-\ln\nu^2)\;.
\end{equation}
In eq. (9), with $\chi=1$, the glueball equation of motion can be dropped and
we are left with the single scalar equation
\begin{equation}
\sigma_0^2D\nu-2B_0\delta\nu=M\rho_s-B_0\delta\frac{(1+\nu^2)}{\nu}
-m_{\omega}^2R_{\omega}\omega_0^2\nu\;.
\end{equation}
Taking care that the energy is consistently obtained we find
\begin{eqnarray}
E&\pj=&\pj\sum_{\alpha}^{\rm occ}\epsilon_{\alpha}(2j_{\alpha}+1)
-2\pi\int_0^{\infty}drr^2\bigg\{M\nu\rho_s+g_{\omega}\omega_0\rho_B
+g_{\rho}b_0\rho_3\nonumber\\
&&+2B_0\delta\ln\nu-m_{\omega}^2R_{\omega}\nu^2\omega_0^2\bigg\}\;.
\end{eqnarray}
Using the low density expansion for nuclear matter and replacing $B_0$ by
$B_0'$ in eqs. (18) and (19) this expression
gives the same result as the complete result of eq. (10). In practice to
saturate nuclear matter correctly a small adjustment of $C_{\omega}^2$ is
needed and $B_0'$ is only approximately given by (16).
\newpage
\centerline{\bf Table 1}
\centerline{Values of the parameters and the derived quantities
for nuclear matter}
\begin{center}
\begin{tabular}{|c|ccccc|ccc|}\hline
Set&$R_{\omega}$&$33\delta$&
${\displaystyle |\epsilon_{\rm vac}|^{\frac{1}{4}}}$&$C_{\omega}^2$
&$C_{\rho}^2$&${\displaystyle \frac{\strut M^*_{\rm sat}}{M}}$
&$K$&$\frac{S}{K}$\\[1mm]
&&&(MeV)&&&&(MeV)&\\ \hline
I&0.0&4&235&156&116&0.71&383&6.5\\
IF&0.0&4&231&152&123&0.72&341&5.6\\
II&0.5&4&255&82.6&127&0.80&356&5.9\\
III&1.0&4&269&51.3&132&0.84&377&6.1\\
IIIF&1.0&4&265&49.2&135&0.85&356&5.7\\
IV&0.0&1&336&153&122&0.72&350&5.8\\
V&0.0&2&281&154&119&0.72&360&6.0\\
VI&0.0&8&194&158&107&0.69&442&7.5\\
VII&0.0&16&160&156&88&0.65&666&10.6\\ \hline
\end{tabular}
\end{center}
\newpage
\centerline{\bf Table 2}
\centerline{Bulk properties of nuclei for various parameter sets}
\begin{center}
\begin{tabular}{|ccc|cc|cc|cc|} \hline
&&&\multicolumn{2}{|c|}{O}&\multicolumn{2}{c|}{Ca}&\multicolumn{2}{c|}{Pb}\\
Set&$m_<$&$\zeta$&$\frac{BE}{A}$&$r_{ch}$&$\frac{BE}{A}$&$r_{ch}$&
$\frac{BE}{A}$&$r_{ch}$\\
&(MeV)&&(MeV)&fm&(MeV)&fm&(MeV)&fm\\ \hline
Experiment&&&7.98&2.73&8.45&3.48&7.86&5.50\\ \hline
I/110&506&1.4&7.35&2.64&7.96&3.41&7.33&5.49\\
IF/110&508&--&7.86&2.62&8.35&3.40&7.54&5.49\\
I/93&598&1.4&9.41&2.52&9.38&3.32&8.04&5.45\\
I/120&464&1.4&6.25&2.72&7.16&3.47&6.92&5.52\\
II/110&598&1.6&10.08&2.49&9.85&3.29&8.22&5.45\\
III/110&664&1.8&10.98&2.44&10.45&3.26&8.48&5.44\\ \hline
\end{tabular}
\end{center}
\newpage
\begin{center}
{\bf Figure Captions}
\end{center}
\noindent Fig. 1.\hspace{.3cm}The binding energy/particle as a function of
density for nuclear matter; density $\rho$ is given as the ratio to the
equilibrium value $\rho_0=0.148$ fm$^{-3}$. The parameter sets used are
indicated and the designation F indicates the frozen glueball model.

\noindent Fig. 2.\hspace{.3cm}The fields $\nu$ and $\chi$ corresponding
to fig. 1.

\noindent Fig. 3.\hspace{.3cm}Comparison of the experimental charge density
\cite{vri} for $^{40}$Ca with theoretical predictions for the parameter
sets indicated.

\noindent Fig. 4\hspace{.3cm}As for fig. 3, but for $^{208}$Pb.

\noindent Fig. 5\hspace{.3cm}As for fig. 3, but for $^{16}$O . The data are
from ref. \cite{sick}.

\noindent Fig. 6.\hspace{.3cm}Occupied and unoccupied neutron levels near the
Fermi energy in $^{208}$Pb. The experimental data are compared with
predictions for the parameter sets indicated.

\noindent Fig. 7.\hspace{.3cm}As for fig. 6, but for protons.

\begin{thebibliography}{999}
\bibitem{mig} J. Schechter, Phys. Rev. {\bf D21} (1980) 3393; A.A. Migdal and
M.A. Shifman, Phys. Lett. {\bf B114} (1982) 445.
\bibitem{eko1} J. Ellis, J.I. Kapusta and K.A. Olive, Nucl. Phys. {\bf B348}
(1991) 345.
\bibitem{glu1} P.J. Ellis, E.K. Heide and S. Rudaz, Phys. Lett.
{\bf B282} (1992) 271.
\bibitem{rho} E.K. Heide, Ph.D. Thesis, unpublished (1991); I. Mishustin,
 J. Bondorf and M. Rho, preprint (1992).
\bibitem{rod} R.G. Rodr\'{\i}guez and J.I. Kapusta, Phys. Rev. {\bf C44}
(1991) 870.
\bibitem{glu2} E.K. Heide, S. Rudaz and P.J. Ellis, Phys. Lett. {\bf B293}
(1992) 259.
\bibitem{fern} R.J. Furnstahl and B.D. Serot, Phys. Rev. {\bf C47} (1993) 2338.
\bibitem{bog} J. Boguta, Phys. Lett. {\bf B120} (1983) 34.
\bibitem{sew} B.D. Serot and J.D. Walecka, Advances in Nuclear Physics,
Vol. 16, ed. J.W. Negele and E. Vogt (Plenum, NY, 1986); B.D. Serot,
Rep. Prog. Phys. {\bf 55} (1992) 1855.
\bibitem{gomm} H. Gomm and J. Schechter, Phys. Lett. {\bf B158} (1985) 449.
\bibitem{crew} R.J. Crewther, Phys. Rev. Lett. {\bf28} (1972) 1421;
M.S. Chanowitz and J. Ellis, Phys. Lett. {\bf B40} (1972) 397;
Phys. Rev. {\bf D7} (1973) 2490; J.C. Collins, A. Duncan and S.D. Joglekar,
Phys. Rev. {\bf D16} (1977) 438.
\bibitem{itep} A.I. Vainshtein, V.I. Zakharov, V.A. Novikov and M.A. Shifman,
Sov. J. Part. Nucl. {\bf13} (1982) 224.
\bibitem{horse} C.J. Horowitz and B.D. Serot, Nucl. Phys. {\bf A368} (1981)
503.
\bibitem{oades} O. Dumbrajs {\it et al.}, Nucl. Phys. {\bf B216} (1983) 277;
J. Hamilton and G.C. Oades, Nucl. Phys. {\bf A424} (1984) 447.
\bibitem{svz} M.A. Shifman, A.I. Vainshtein and V.I. Zakharov, Nucl. Phys.
{\bf B147} (1979) 448.
\bibitem{bag} A. Chodos, R.L. Jaffe, K. Johnson, C.B. Thorn and V.F. Weisskopf,
Phys. Rev. {\bf D9} (1974) 3471; T. DeGrand, R.L. Jaffe, K. Johnson
and J. Kiskis, Phys. Rev. {\bf D12} (1975) 2060.
\bibitem{pear} J.M. Pearson, Phys. Lett. {\bf B271} (1991) 12.
\bibitem{skpap} S. Rudaz, P.J. Ellis, E.K. Heide and M. Prakash, Phys. Lett.
{\bf B285} (1992) 183.
\bibitem{shif} M.A. Shifman, Z. Phys. {\bf C9} (1981) 347; P. Pascual and
R. Tarrach, Phys. Lett. {\bf B113} (1982) 495.
\bibitem{linsig} M. Gell-Mann and M. L\'evy, Nuovo Cimento {\bf16} (1960) 705.
\bibitem{gas} S. Gasiorowicz and D.A. Geffen, Rev. Mod. Phys.
{\bf41} (1969) 531.
\bibitem{neg} J.W. Negele, Phys. Rev. {\bf C1} (1970) 1260.
\bibitem{tas} L.J. Tassie and F.C. Barker, Phys. Rev. {\bf111} (1958) 940.
\bibitem{shep} C.E. Price, J.R. Shepard and J.A. McNeil, Phys. Rev. {\bf C41}
(1990) 1234; {\it ibid.} {\bf C42} (1990) 247.
\bibitem{vri} H. de Vries, C.W. de Jager and C. de Vries, At. Data and
Nucl. Data Tables, {\bf36} (1987) 495.
\bibitem{sick} I. Sick and J.S. McCarthy, Nucl. Phys. {\bf A150} (1970) 631.
\bibitem{fern2} R.J. Furnstahl, C.E. Price and G.E. Walker, Phys. Rev. {\bf
C36}
(1987) 2590.
\bibitem{fs2} R.J. Furnstahl and B.D. Serot, Indiana University preprint,
\# IU/NTC 93--15.
\end{thebibliography}
\end{document}